\documentclass[reprint,superscriptaddress,amsmath,amssymb,aps]{revtex4-1}

\usepackage{color}
\usepackage{graphicx}
\usepackage{dcolumn}
\usepackage{bm}
\usepackage{hyperref}
\usepackage{braket}
\usepackage{mathtools}
\usepackage{float}

\newcommand{\calC}{\mathcal{C}}
\newcommand{\calE}{\mathcal{E}}

\newcommand{\re}{\mathrm{Re}}

\begin{document}
\preprint{APS/123-QED}

\title{Current-induced second harmonic generation in  inversion-symmetric Dirac and Weyl semimetals}

\author{Kazuaki Takasan}
\email{takasan@berkeley.edu}
\affiliation{%
Department of Physics, University of California, Berkeley, California 94720, USA}%

\author{Takahiro Morimoto}
\affiliation{%
Department of Applied Physics, The University of Tokyo, Hongo, Tokyo, 113-8656, Japan
}%
\affiliation{%
JST, PRESTO, Kawaguchi, Saitama, 332-0012, Japan
}%

\author{Joseph Orenstein}
\affiliation{%
Department of Physics, University of California, Berkeley, California 94720, USA}%
\affiliation{%
Materials Science Division, Lawrence Berkeley National Laboratory, Berkeley,
California 94720, USA}%

\author{Joel E. Moore}
\affiliation{%
Department of Physics, University of California, Berkeley, California 94720, USA}%
\affiliation{%
Materials Science Division, Lawrence Berkeley National Laboratory, Berkeley,
California 94720, USA}%

\date{\today}
\begin{abstract}
Second harmonic generation (SHG) is a fundamental nonlinear optical phenomenon widely used both for experimental probes of materials and for application to optical devices. Even-order nonlinear optical responses including SHG generally require breaking of inversion symmetry, and thus have been utilized to study noncentrosymmetric materials. Here, we study theoretically the SHG in inversion-symmetric Dirac and Weyl semimetals under a DC current which breaks the inversion symmetry by creating a nonequilibrium steady state. Based on analytic and numerical calculations, we find that Dirac and Weyl semimetals exhibit strong SHG upon application of finite current. Our experimental estimation for a Dirac semimetal Cd$_3$As$_2$ and a magnetic Weyl semimetal Co$_3$Sn$_2$S$_2$ suggests that the induced susceptibility $\chi^{(2)}$ for practical applied current densities can reach $10^5~\mathrm{pm}\cdot\mathrm{V}^{-1}$ with mid-IR or far-IR light. This value is 10$^2$-10$^4$ times larger than those of typical nonlinear optical materials.
We also discuss experimental approaches to observe the current-induced SHG and comment on current-induced SHG in other topological semimetals in connection with recent experiments.
\end{abstract}

\maketitle 

\textit{Introduction.---} 
Intense light incident on materials induces various nonlinear optical responses (NLORs) reflecting the details of material properties~\cite{Bloembergen_book, Boyd_book}. The study of NLORs remains an important topics in condensed matter studies since NLORs not only give a rich information of symmetry information about materials but also yield useful optical devices. In recent years, a close relationship between the NLORs and the notion of band geometry has been revealed~\cite{moore-orenstein, Deyo2009, Sodemann2015, sipe, young-rappe, Morimoto2016Floquet}. In particular, three-dimensional (3D) topological materials can support novel NLORs~\cite{hosur11, chan17, Wu2017, Juan2017, Golub2018, Patankar2018}.  Among these, inversion-symmetry-broken topological semimetals (SMs) are attracting keen attention as recent optical measurements of TaAs, which is an inversion-symmetry-broken Weyl semimetal (WSM), reported strong second harmonic generation (SHG) with signal 100 times larger than a typical value in GaAs~\cite{Wu2017, Patankar2018}, and other strong nonlinear optical properties as well~\cite{Osterhoudt2019, Sirica2019}. 
From the theoretical side, various interesting nonlinear optical phenomena have been proposed~\cite{hosur11, chan17, Juan2017}, including a quantization of the circular photogalvanic effect that originates from the topological properties of WSMs~\cite{Juan2017,flicker18}. 

On the other hand, there are various topological SMs preserving inversion symmetry which are also intensively studied. One example is topological Dirac semimetals (DSMs), such as Cd$_3$As$_2$~\cite{Rosenberg1959, Wang2013, Crassee2018} or Na$_3$Bi~\cite{Wang2012}, where the Dirac point is protected by crystalline symmetry. 
The other example is inversion-symmetric magnetic WSMs, such as Co$_3$Sn$_2$S$_2$~\cite{Liu2019} or Mn$_3$Sn~\cite{Kuroda2017}, where the time-reversal symmetry is broken instead of the inversion symmetry. In these materials, odd-order NLORs are only allowed, where the dominant effect is the third-order NLOR. 
However, once the inversion symmetry is broken by applying a suitable perturbation, inversion-symmetric materials can also exhibit even order NLORs.

\begin{figure}[t]
\includegraphics[width=9cm]{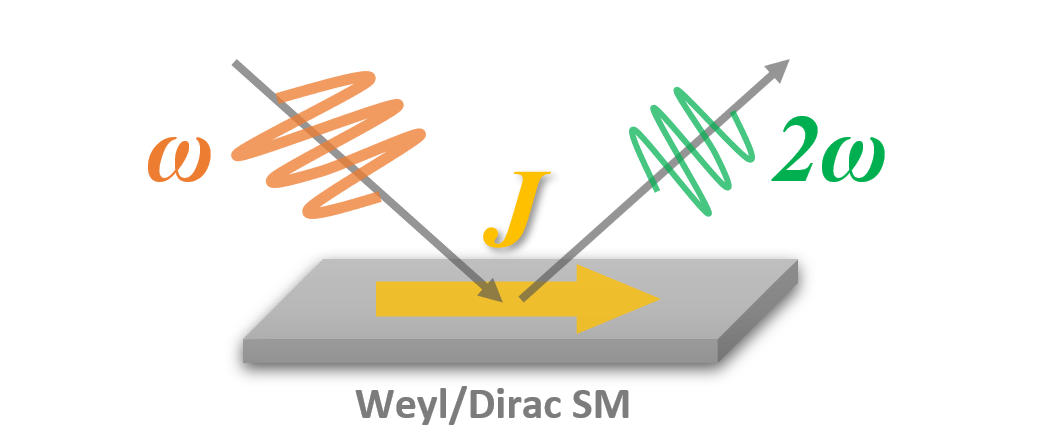}
\caption{
Concept of currrent-induced second harmonic generation.  Dirac/Weyl semimetals (SMs) in a nonequilibrium state carrying finite current show induced second-harmonic generation (SHG), i.e., probe light with frequency $\omega$ is converted into outgoing light with frequency $2\omega$. 
}
\label{Fig:Setup}
\end{figure}

Motivated by this idea and by general interest in the creation of nonequilibrium states with new or amplified responses, we investigate the creation of second-order NLORs in inversion-symmetric Dirac/Weyl SMs. For the inversion-symmetry breaking perturbation, we consider DC electric field which makes the electron distribution asymmetric in momentum space and induces finite current, resulting in broken inversion symmetry. In this study, we focus on SHG, which is a phenomenon that injected light with frequency $\omega$ is converted into light with doubled frequency $2\omega$ as schematically shown in Fig.~\ref{Fig:Setup}. SHG from current-driven materials has been called current-induced SHG (CISHG) and studied theoretically~\cite{Khurgin1995, Wu2012, Cheng2014} and experimentally in several materials, such as Si~\cite{Aktsipetrov2009}, GaAs~\cite{Ruzicka2012},  graphene~\cite{Bykov2012,  An2013} and superconducting NbN~\cite{Nakamura2020}. This device geometry is similar to what is used to measure photoconductivity in inversion-symmetric insulators, where an applied DC electric field leads to a current pulse under illumination. In metals, this DC field will already produce some current, so the most visible optical consequence of the field-induced symmetry breaking is now CISHG.

Yet, CISHG in topological materials, especially Dirac/Weyl SMs, has not been explored so far.  Here, we study CISHG in Dirac/Weyl SMs by taking two complementary approaches. One is analytic calculation with ideal Weyl (Dirac) Hamiltonians and the other is numerical, based on tight-binding models. The results of both approaches are consistent and show that inversion-symmetric Dirac/Weyl SMs support a divergently large CISHG when the Fermi level is located near the Dirac/Weyl points. Based on our results, we estimate the order of the nonlinear susceptibility $\chi^{(2)}$, characterizing the strength of the CISHG. Considering the realistic parameters corresponding to the materials, a Dirac SM, Cd$_3$As$_2$, and a Weyl SM, Co$_3$Sn$_2$S$_2$, and find that it can reach $10^5~\mathrm{pm}\cdot\mathrm{V}^{-1}$ for practical applied current densities. These values are 10$^2$-10$^4$ times larger than those of typical nonlinear materials~\cite{Wu2017, Bergfeld2003, Haislmaier2013}. 
We also address the experimental methods to observe the CISHG and the possibility of CISHG in other topological SMs. 

\textit{Methods.---} SHG is characterized by the response tensor $\sigma^{abc}_\mathrm{SHG}$ defined via $j^a (2 \omega)=\sigma^{abc}_\mathrm{SHG}(\omega) E^b (\omega) E^c (\omega)$ where $j^a(2\omega)$ [$E^a (\omega)$] is the Fourier component of the time-dependent current $j^a(t)$ [electric field $E^a(t)$] proportional to $e^{2i\omega t}$ [$e^{i\omega t}$]. The indices $a, b, c$ run over $\{x, y, z\}$ and the sum over repeated indices is implied throughout this paper. From the standard time-dependent perturbation theory~\cite{Moss1990, Ghahramani1991, Yang2017}, we have the following expression for the SHG response tensor  
\begin{align}
    \sigma^{abc}_\mathrm{SHG}(\omega) = \sigma^{abc}_\mathrm{2p, I}(\omega)+\sigma^{abc}_\mathrm{2p, II}(\omega) + \sigma^{abc}_\mathrm{1p, I}(\omega)+\sigma^{abc}_\mathrm{1p, II}(\omega), \label{eq:SHG}
\end{align}
where
\begin{align}
    \sigma^{abc}_\mathrm{2p, I}(\omega)&=\frac{e^3}{2\hbar^2 \omega^2} \int [d \bm k]  v^a_{mn} w^{bc}_{nm} f_{mn} R_\gamma (2\omega-\omega_{nm}) \label{eq:SHG2pI} \\
    \sigma^{abc}_\mathrm{2p, II}(\omega)&=\frac{e^3}{2\hbar^2 \omega^2}  \int [d \bm k] \frac{2 v^a_{mn} \{v^b_{np} v^c_{pm}\}}{\omega_{mp}+\omega_{np}}f_{mn} R_\gamma (2\omega-\omega_{nm})
    \label{eq:SHG2pII}\\
    \sigma^{abc}_\mathrm{1p, I}(\omega)&=\frac{e^3}{2\hbar^2 \omega^2}  \int [d \bm k] (w^{ab}_{mn} v^{c}_{nm}+w^{ac}_{mn} v^{b}_{nm}) \nonumber \\ & \qquad \qquad \qquad \quad \times f_{mn} R_\gamma (\omega-\omega_{nm}) \label{eq:SHG1pI}\\
    \sigma^{abc}_\mathrm{1p, II}(\omega)&=\frac{e^3}{2\hbar^2 \omega^2} \int [d \bm k] \frac{ v^{a}_{mn} \{v^b_{np} v^c_{pm}\}}{\omega_{pm}+\omega_{pn}} \nonumber \\ & \qquad  \times \left\{f_{mp} R_\gamma (\omega-\omega_{pm})  -f_{np} R_\gamma (\omega-\omega_{np})\right\}, \label{eq:SHG1pII}
\end{align}
with $v^a=(1/\hbar)\partial_{k_a}H$, $w^{ab}=(1/\hbar)\partial_{k_a}\partial_{k_b}H$, $\{v^b_{np} v^c_{pm}\}=v^b_{np} v^c_{pm}+v^c_{np} v^b_{pm}$, $f_{mn}=f_m -f_n$, $f_n = f(\varepsilon_n)$, $\omega_{mn}=(\varepsilon_m -\varepsilon_n)/\hbar$ and $R_\gamma (x)=1/(x-i\gamma)$ where the integration $\int [d\bm k]\equiv \int d k_x d k_y d k_z /(2\pi)^3$ is performed over the entire Brillouin zone. Here, $\varepsilon_n=\varepsilon_n(\bm k)$ represents the $n$-th band of the Hamiltonian $H=H (\bm k)$ (the implicit sum over repeated indices is also taken for the band indices $m, n$ and $p$) and $f(\varepsilon)$ is a distribution function of electrons. In equilibrium $f(\varepsilon)=f^{(0)}(\varepsilon)\equiv(1+e^{\beta \varepsilon})^{-1}$ which is the Fermi distribution function with inverse temperature $\beta$. The subscript 2p (1p) in Eq.~(\ref{eq:SHG}) denotes the contribution of two-photon (one-photon) resonance.

\begin{figure}[t]
\includegraphics[width=8.5cm]{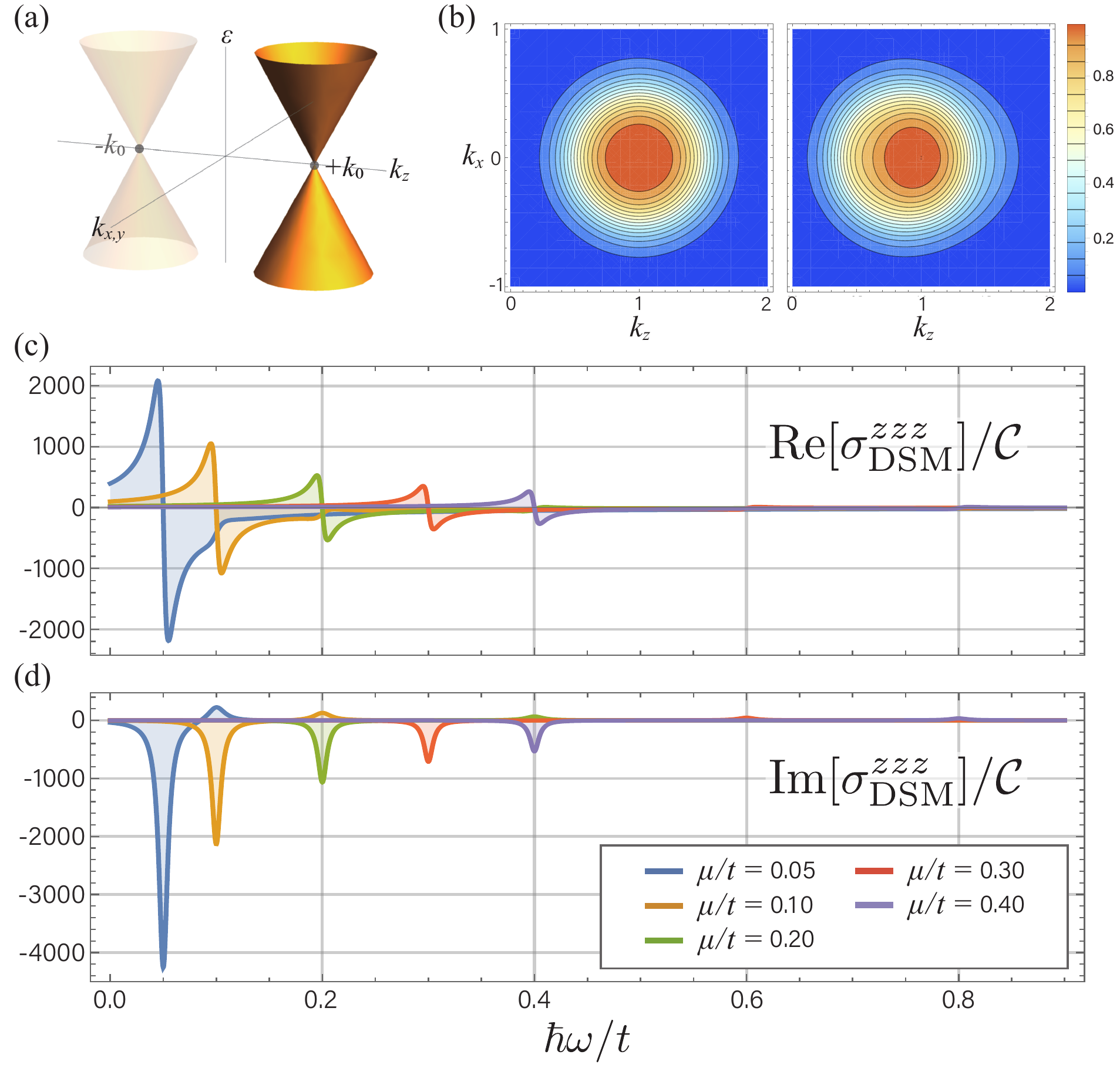}
\caption{
(a) Energy dispersion of the Weyl Hamiltonian [Eq.~(\ref{eq:WeylHam})] whose gapless point is located at $\pm \bm k_0=(0,0, \pm k_0)$. (b) Equilibrium distribution function $f^{(0)}(\varepsilon(\bm k))$ (left) and nonequilibrium distribution function $f(\varepsilon(\bm k))$ up to the order of $\calE_a^2$ (right) in momentum space ($k_y$ is fixed to zero). Here, $\varepsilon(\bm k)$ denotes the larger eigenvalue of the Weyl Hamiltonian and the parameters are set as $t=1$, $a=1$, $\mu=0.5$, $k_0=1.0$, $\beta=10$, and $(\calE_x, \calE_y, \calE_z) =(0, 0, 0.1)$. (c, d) Real and imaginary part of the CISHG response tensor of Dirac semimetals $\sigma^{zzz}_\mathrm{DSM}(\omega)$ calculated with the Weyl Hamiltonian. The values of the vertical axes are normalized by a constant $\calC=(e\tau Ea/\hbar)\cdot(\hbar/t)\cdot(e^3/h^2)$. Here, we set $\gamma=0.01(t/\hbar)$.}
\label{Fig:Ana}
\end{figure}

\begin{figure*}[t]
\includegraphics[width=17.8cm]{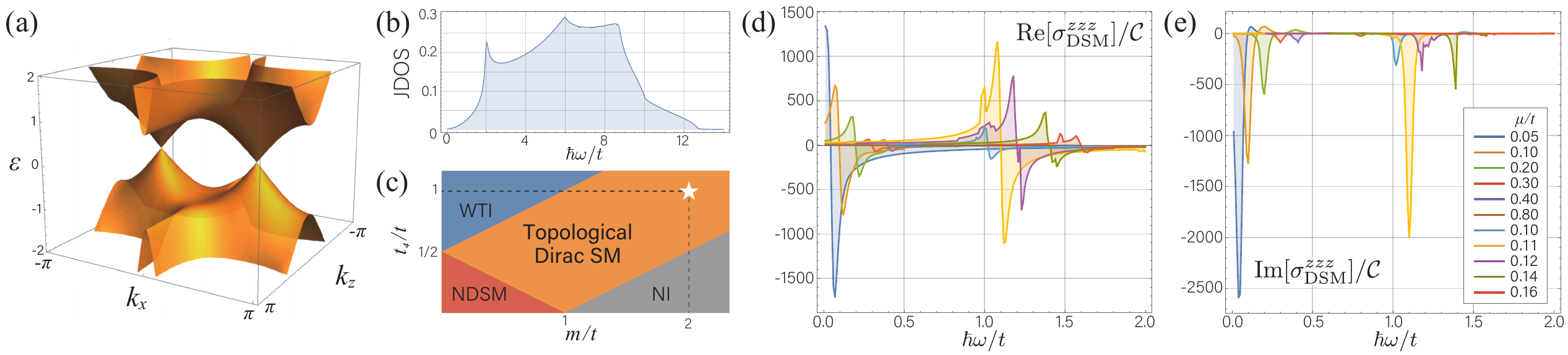}
\caption{
(a) Energy dispersion, (b) joint density of states (JDOS), and (c) topological phase diagram of the tight-binding model [Eq.~(\ref{eq:TBmodel})]. In the topological phase diagram, NDSM, WTI and NI denote normal (i.e. topologically trivial) Dirac semimetal, weak topological insulator, and normal insulator, respectively. The white star symbol represents the parameter that we use in our calculation. (d, e) Real and imaginary part of the CISHG response tensor of Dirac semimetals $\sigma^{zzz}_\mathrm{DSM}(\omega)$ calculated with the tight-binding Hamiltonian. The values in the vertical axes are normalized by a constant $\calC=(e\tau E_z a/\hbar)\cdot(\hbar/t)\cdot(e^3/h^2)$. The parameters that we used are $t=1.0$, $t_1=1.0$, $t_2=2.0$, $t_3=1.0$, $t_4=1.0$, $m=2.0$, $\beta=100$ and $\gamma=0.01(t/\hbar)$. 
}
\label{Fig:TB}
\end{figure*}

To calculate the response tensor of CISHG, we need the distribution function of a nonequilibrium steady state (NESS) carrying finite current. To obtain it, we use the Boltzmann equation with the relaxation time approximation under a static electric field $\bm E_\mathrm{DC}$, which is $-(e \bm E_\mathrm{DC}/\hbar)\cdot(\partial f_n/\partial \bm k)=-(f_n-f^{(0)}_n)/\tau$ where $\tau$ denotes the relaxation time~\cite{Yu2o14, Sodemann2015, Morimoto2016DiracWeyl, Yasuda2016}. Solving this equation recursively, we obtain the distribution function for NESS as
\begin{align}
    f_n = f^{(0)}_n + \calE_a \frac{\partial f^{(0)}_n}{\partial k_a} +  \calE_a \calE_b  \frac{\partial^2 f^{(0)}_n}{\partial k_a \partial k_b} + \cdots \label{eq:NESSdist}
\end{align}
with $\calE_a = e \tau E^a_\mathrm{DC}/\hbar$. We use $f_n$ in Eq.(\ref{eq:SHG2pI})-(\ref{eq:SHG1pII}) to calculate the current-induced SHG. The example of $f_n$ under the electric field in $z$-direction is shown in the right panel of Fig.~\ref{Fig:Ana}~(b) and the equilibrium distribution $f_n^{(0)}$ is also shown in the left panel of Fig.~\ref{Fig:Ana}~(b) for reference. From these figures, we can see the distribution function is deformed and asymmetric in the $k_z$-direction under the electric field. 

\textit{Analytic results with Weyl Hamiltonian. ---}
To study the CISHG in inversion-symmetric Dirac/Weyl SMs, we take two complementary approaches. One approach is based on a simple Weyl Hamiltonian
\begin{align}
    H_\mathrm{Weyl}&= \chi v (\bm p - \bm p_0) \cdot \bm \sigma - \mu \sigma_0 \label{eq:WeylHam}
\end{align}
where $\sigma_{x,y,z} (\sigma_0)$ represents Pauli matrices ($2 \times 2$ identity matrix), $\bm \sigma=(\sigma^x, \sigma^y, \sigma^z)$, $v=ta/\hbar~$ ($t$ and $a$ correspond the hopping amplitude and the lattice constant respectively), $~\bm p = \hbar \bm k = \hbar (k_x, k_y, k_z)$, and $\bm p_0 = \hbar \bm k_0$. The Hamiltonian $H_\mathrm{Weyl}$ represents a single Weyl ($\chi=+1$) or anti-Weyl ($\chi=-1$) node located at $\bm k = \bm k_0$~[The band structure is shown in Fig.~(\ref{Fig:Ana})~(a)]. This Hamiltonian is very simple, but the low-energy physics of Dirac/Weyl SMs are well-described by this Hamiltonian~\cite{Armitage2018}. WSMs have pairs of Weyl and anti-Weyl nodes in the band structure and they locate at different points. On the other hand, DSMs support Weyl and anti-Weyl nodes at the same point, which is called a Dirac node. In the following, we start from calculation of the SHG of a single (anti-)Weyl node and then sum up contributions from all the Weyl nodes~\cite{FN2}. For simplicity, we assume that the electric fields are applied in the $z$-direction, i.e. $\bm E_\mathrm{DC}=(0, 0, E_z)$.

By a straightforward calculation with the Weyl Hamiltonian (\ref{eq:WeylHam}) shown in Supplemental Material, we can evaluate Eq.~(\ref{eq:SHG}) analytically at zero temperature. 
Considering the symmetry, the independent non-zero components are only $zzz$-, $zxx$- and $xzx$-components~\cite{Bloembergen_book, Boyd_book, FNB}. The $zzz$-component of the response tensor from a single Weyl node is given as 
\begin{align}
    \sigma^{zzz}_\mathrm{single}(\omega) &= \frac{e\tau E_z a}{\hbar} \frac{\hbar}{t} \frac{e^3}{h^2} \left\{- \frac{4}{15} F_\mathrm{2p}(\omega)+\frac{1}{30}F_\mathrm{1p}(\omega)\right\}, \label{eq:SHGWeylHamTot}
\end{align}
with
\begin{align}
    F_\mathrm{2p}(\omega)=\frac{t}{|\mu|}\frac{t/\hbar}{\omega - |\mu/\hbar| - i \gamma/2}, \label{eq:SHGWeylHam2p}\\
    F_\mathrm{1p}(\omega)=
    \frac{t}{|\mu|}\frac{t/\hbar}{\omega - 2|\mu/\hbar| - i \gamma}. \label{eq:SHGWeylHam1p}
\end{align}
The other independent components are given as $\sigma^{zxx}_\mathrm{single}(\omega)= \calC \{- (2/15) F_\mathrm{2p}(\omega)+ (3/10)F_\mathrm{1p}(\omega)\}$ and $\sigma^{xzx}_\mathrm{single}(\omega)=\calC\{(1/5) F_\mathrm{2p}(\omega)-(7/60)F_\mathrm{1p}(\omega)\}$ with $\calC=(e\tau E_z a/\hbar)\cdot(\hbar/t)\cdot(e^3/h^2)$. The differences between the components are only numerical factors and their qualitative behaviors are same. Thus, we focus on the $zzz$-component below.

Using these results, we can obtain the CISHG response tensor for Dirac/Weyl SMs. Since the above results do not depend on the position and the chirality of the Weyl nodes, we can calculate the response tensor of Dirac/Weyl SMs simply by multiplying the number of Weyl nodes considering the degeneracy. Therefore, assuming that Weyl (Dirac) SMs have two Weyl (Dirac) nodes~\cite{FN3}, the response tensors for Weyl and Dirac SMs are $\sigma^{zzz}_\mathrm{WSM}(\omega)=2\sigma^{zzz}_\mathrm{single}(\omega)$ and $\sigma^{zzz}_\mathrm{DSM}(\omega)=4\sigma^{zzz}_\mathrm{single}(\omega)$ respectively. The real and imaginary part of $\re[\sigma^{zzz}_\mathrm{DSM}(\omega)]$ are shown in Figs.~\ref{Fig:Ana}~(c) and (d). From these figures and Eq.~(\ref{eq:SHGWeylHamTot})-(\ref{eq:SHGWeylHam1p}), we can see that the SHG spectra have a large peak around $\hbar \omega=\mu$ (two-photon resonance) and a small peak around $\hbar \omega=2\mu$ (one-photon resonance). The height (weight) of the two peaks is proportional to $1/\mu$ for $\mu \to 0$, leading to diverging enhancement. We note that the power of $1/\mu$ is different from that in the graphene case discussed in the previous study~\cite{Cheng2014} because of the different dimensionality~\cite{FN1}. Our results suggest that Dirac/Weyl SMs where the Fermi level is near the Dirac/Weyl points can show strong SHG~\cite{FNA}. 

\textit{Numerical results with  tight-binding models.---}
Let us move on to the numerical calculation with a tight-binding Hamiltonian describing DSMs. This approach is closer to real materials than the previous approach because we consider multiple (in this model, four) bands and take the nonlinearity and periodicity of the band structure into account. We use the following tight-binding model
\begin{align}
    H_\mathrm{TB}(\bm k) &= f_1(\bm k)
    \sigma_z \tau_x 
    +f_2(\bm k) \sigma_0 \tau_y \nonumber \\
    &\quad + f_3(\bm k) \sigma_x \tau_x + f_4(\bm k) \sigma_y \tau_x + f_5(\bm k)\sigma_0 \tau_3, \label{eq:TBmodel}
\end{align}
with
\begin{align}
f_1(\bm k)&=t_1 \sin(a k_x), f_2(\bm k)= - t_1 \sin (a k_y),\\
f_3(\bm k)&=(t_2 + t_3) [\cos(a k_y)-\cos(a k_x)] \sin(a k_z), \\
f_4(\bm k)&=- (t_2 - t_3) \sin(a k_x) \sin(a k_y) \sin (a k_z),  \\
f_5(\bm k)&=m-t_4\{\cos(ak_x)+\cos(ak_y)\}-t\cos(ak_z),
\end{align}
introduced in Ref.~\cite{Yang2014}. As shown in the topological phase diagram [Fig.~\ref{Fig:TB}~(c)], this model hosts several topological phases. In particular, the topological DSM phase is realized in a wide range of parameters. In this phase, the energy dispersion has a pair of Dirac points located on $k_z$-axis as shown in Fig.~\ref{Fig:TB}~(a). These Dirac cones are protected by the $C_4$ rotational symmetry and topologically robust, which is also the case in the typical topological DSM material, Cd$_3$As$_2$~\cite{Yang2014}. 

Using this model, we calculate the SHG response tensor $\sigma^{zzz}_\mathrm{DSM}(\omega)$ under the $z$-directed electric field $\bm E_\mathrm{DC}=(0, 0, E_z)$ at finite temperature~\cite{FN4}. By numerical calculations, we obtain the SHG spectra shown in Figs.~\ref{Fig:TB}~(d) and (e). First, we can find strong peaks around $\mu \sim 0$ and they show a divergent behavior as $\mu \to 0$. These signatures are consistent with our analytic results shown in Figs.~(\ref{Fig:Ana})~(c) and (d)~\cite{FN9}. These findings strongly suggest that Dirac/Weyl SMs generally support large CISHG. The other feature in the spectra is the appearance of a large peak at $\omega \sim \mu/\hbar$ when $\mu/t \sim 1.0$-$1.4$. This behavior reflects the van Hove singularity at $\mu/t=1.0$. Indeed, the joint density of states (JDOS)~\cite{FN8} shows a singularity at $\hbar\omega/t \sim 2.0$ as shown in Fig.~\ref{Fig:TB}~(b)~\cite{FN7}.

\begin{figure}[t]
\includegraphics[width=8.5cm]{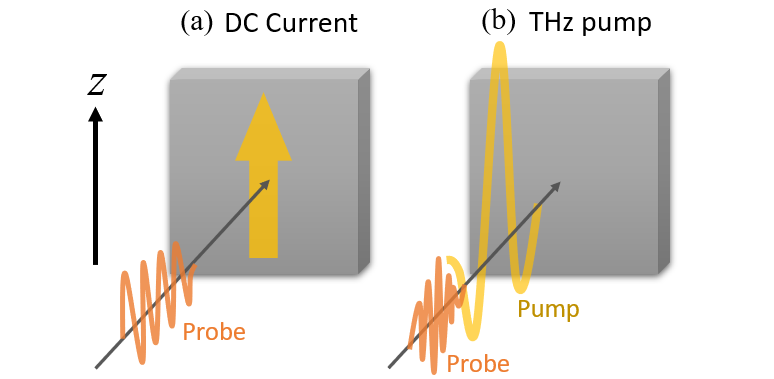}
\caption{
Two experimental approaches to observation of CISHG. (a) Standard SHG measurement with applying a DC bias voltage to induce a current. (b) Pump-probe-type SHG measurement. The pump pulse should be at low enough frequency compared to interband excitations that the induced state is the same as that created by a DC voltage, which for most materials extends up to the terahertz (THz) range. 
}
\label{Fig:ExpMethod}
\end{figure}

In addition to Dirac SMs, we also carried out a tight-binding calculation for Weyl SMs (See Supplemental Material). We study a two-band tight-binding model describing Weyl SMs and obtain qualitatively similar results to those of Dirac SMs. Therefore, strong CISHG in Weyl SMs is also supported by both analytic and numerical calculations. 

\textit{Discussion.---} 
Our calculation suggests that Dirac and Weyl SMs show very strong CISHG. 
To connect these results with experiments, we estimate the strength of CISHG.  First of all, we need to specify an experimental setup to give an estimate because the achievable electric field depends on the type of experiment. We propose two kinds of experimental setup shown in Fig.~\ref{Fig:ExpMethod}. One is a standard SHG measurement under a DC bias voltage. The other is a THz pump SHG measurement, where the pump frequency is low enough to be seen as a static field. The former approach is static and thus should be easier than the other one, which is time-resolved. On the other hand, the latter approach is advantageous for applying a strong electric field because very strong THz fields such as 1-80 MV/cm has been achieved~\cite{Fulop2020}. 

Next, we estimate the strength of electric fields inside the material. For the DC bias case, the experimental control parameter is current density rather than field strength. Following Ohm's law, the internal electric field $E_\mathrm{in}$ is given as $E_\mathrm{in}=j/\sigma$, where $j$ and $\sigma$ are the current density and the conductivity, respectively. For the THz pump case, we have to take into account the mismatch of impedance.
The internal electric field is represented as $E_\mathrm{in}=2E_\mathrm{ext}/(n+1)$ with the external pump field $E_\mathrm{ext}$ and the refractive index $n$. In the THz regime, the refractive index is given as $n=\sqrt{\sigma/(i \Omega \varepsilon_0)}$ where $\varepsilon_0$ is the vacuum permittivity and $\Omega$ is the pump field frequency. To be specific, we consider two materials: a Dirac semimetal, Cd$_3$As$_2$ and a Weyl semimetal, Co$_3$Sn$_2$S$_2$. Using the low temperature conductivity of these materials~\cite{FN5} and assuming $j \sim 10^7~\mathrm{A/m^2}$, $E_\mathrm{ext} \sim 400~\mathrm{kV/m}$ and $\Omega \sim 0.5$~THz as typical values, we obtain $E_\mathrm{in}\sim 4.4~(30)~\mathrm{V/m}$ for the DC bias case and $E_\mathrm{in} \sim 2~(4)~\mathrm{kV/cm}$ for the THz pump case in Cd$_3$As$_2$ (Co$_3$Sn$_3$S$_2$).

To estimate the strength of CISHG, we evaluate the nonlinear susceptibility $\chi^{zzz} = \sigma^{zzz}/(2i\omega\varepsilon_0)$. The susceptibility takes the largest value when the probe frequency is resonant to the Fermi energy, i.e. $\omega \sim \mu/\hbar$, and we consider this resonant case below. The Fermi energy of Cd$_3$As$_2$ (Co$_3$Sn$_3$S$_2$) is typically 100~(50)~meV~\cite{Liu2014, Crassee2018, Liu2019} and thus the frequency of probe light is 24~(12)~THz, which corresponds to the wavelength 12.5~(25)~$\mu$m in the mid-IR (far-IR) regime. Using the formula~(\ref{eq:SHGWeylHamTot}) with the parameters for Cd$_3$As$_2$ (Co$_3$Sn$_2$S$_2$)~\cite{FN6}, we obtain $|\re \chi^{zzz}|\sim1.2\times 10^3$~($1.1\times 10^2$)~pm$\cdot$V$^{-1}$ in the DC bias case and $|\re \chi^{zzz}|\sim5.5\times 10^5$~($1.1\times 10^5$)~pm$\cdot$V$^{-1}$ in the THz pump case. Compared to the typical value of the susceptibility of SHG, such as $3.6 \times 10^{3}$~pm$\cdot$V$^{-1}$ of TaAs (the fundamental wavelength $\lambda=$~800 nm)~\cite{Wu2017}, $3.5 \times 10^{2}$~pm$\cdot$V$^{-1}$ of GaAs ($\lambda=$~810~nm)~\cite{Bergfeld2003}, 15-19~pm$\cdot$V$^{-1}$ of BiFeO$_3$ ($\lambda=$~1.55~$\mu$m)~\cite{Haislmaier2013}, the values evaluated above are very large and suggest that Cd$_3$As$_2$ and Co$_3$Sn$_2$S$_2$ are promising candidates showing very strong CISHG. For the DC bias case, the response is relatively small because the internal electric fields are small, but CISHG of Cd$_3$As$_2$ can be comparable to SHG of TaAs due to its longer relaxation time. Remarkably, the responses of both Cd$_3$As$_2$ and Co$_3$Sn$_2$S$_2$ in the THz pump case is 10$^2$ times larger than that of TaAs, which has the largest $\chi^{(2)}$, and 10$^4$ times larger than that of BiFeO$_3$.

Next, we mention other frequency regimes. In the THz regime, the response is expected to be much larger than the mid-IR regime since the CISHG becomes divergently large with $\mu \to 0$. The analytic result~[Eq.~(\ref{eq:SHGWeylHamTot})] indicates that the resonant response at $\hbar \omega/t=1$-$10$~meV (i.e. 0.24-2.4~THz) is roughly 10-100 times larger than that at $\hbar \omega/t=100$~meV. This enhancement is expected to be realized by changing the doping level. For example, the sample of Cd$_3$As$_3$ with $E_F\sim 0$ (i.e. at the Dirac point) has been fabricated as used in Ref.~\cite{Liu2014}. In the higher frequency regime, such as near-IR or visible regime, the effect of the Dirac cones becomes smaller, while van Hove singularity points due to merging of the Dirac cones give rise to a large CISHG response. This contribution can be comparable to the contribution of Dirac points as shown in Figs.~\ref{Fig:TB}~(d)~and~(e). 

We comment on the possibility of the CISHG in other topological SMs. Since our analytic results are only based on the simple Weyl point Hamiltonian without any assumption about symmetry, the similar CISHG can occur even in inversion-symmetry-broken Weyl SMs, such as TaAs. Such materials are expected to show a large CISHG in addition to the original SHG, and these two contributions are separable via changing the applied electric field.  Very recent  experiments~\cite{Sirica2020} suggest that indeed the CISHG component is detectable in TaAs using an optically pumped current, which is found to change the symmetry of SHG in the plane perpendicular to that material's polar axis; our model predicts that the signal induced in Cd$_3$As$_2$ should be much stronger because its relaxation time is at least an order of magnitude longer.  Furthermore, our tight-binding calculation suggests that the van Hove singularities can be an origin of a large CISHG while they are not divergent like the CISHG from Weyl nodes. Thus, since topological nodal SMs have van Hove singularities protected by its topology, they are also expected to be candidate materials showing strong CISHG.

In this paper, we have shown that Dirac/Weyl SMs with inversion symmetry show very strong CISHG, and inversion-breaking topological SMs may also be expected to show strong CISHG on top of the zero-current ordinary SHG. These results suggest that topological SMs have value as a nonlinear optical material whose SHG intensity is controllable from zero to very large value by electric current. Moreover, the SHG is also controlled by changing the direction of the current. This high degree of control can provide a new route to realize switchable nonlinear optical devices.

\begin{acknowledgments}
We thank Daniel E. Parker for valuable discussions. This work is supported by the Quantum Materials Program (JWO, JEM) and a Simons Investigatorship (JEM).
KT thanks JSPS for support from Overseas Research Fellowship.
TM was supported by JST PRESTO (JPMJPR19L9) and JST CREST (JPMJCR19T3).
\end{acknowledgments}

\bibliographystyle{apsrev4-1}
\bibliography{ref.bib}

\begin{thebibliography}{57}
\bibitem [{FN2()}]{FN2}%
  \BibitemOpen
  \href@noop {} {}\bibinfo {note} {There are cases, such as the chiral magnetic
  effect, in which treating Weyl nodes in isolation of the band structure
  between them is dangerous. In this case, the issue does not arise as we are
  perturbing the ground state occupations only near the Weyl points and the
  ground-state response is zero by symmetry}\BibitemShut {NoStop}%
\bibitem [{FNB()}]{FNB}%
  \BibitemOpen
  \href@noop {} {}\bibinfo {note} {From the symmetry,
  $\sigma^{zxx}_\mathrm{single}=\sigma^{zyy}_\mathrm{single}$ and
  $\sigma^{xzx}_\mathrm{single}=\sigma^{xxz}_\mathrm{single}=\sigma^{yzy}_\mathrm{single}=\sigma^{yyz}_\mathrm{single}$
  and the other components are zero.}\BibitemShut {Stop}%
\bibitem [{FN3()}]{FN3}%
  \BibitemOpen
  \href@noop {} {}\bibinfo {note} {A single Dirac node is the simplest
  assumption, but we consider two Dirac nodes here to compare this result with
  our tight-binding calculation respecting the features of a Dirac SM material
  Cd$_3$As$_2$}\BibitemShut {NoStop}%
\bibitem [{FNA()}]{FNA}%
  \BibitemOpen
  \href@noop {} {}\bibinfo {note} {The other important point is that the
  strength of SHG does not depend on both the position and the chirality of the
  Weyl node, and thus the Dirac/Weyl SMs can show the strong CISHG as long as
  there exist Dirac/Weyl nodes.}\BibitemShut {Stop}%
\bibitem [{FN4()}]{FN4}%
  \BibitemOpen
  \href@noop {} {}\bibinfo {note} {We consider finite temperature only for
  numerical purpose. The SHG is stronger at lower temperature}\BibitemShut
  {NoStop}%
\bibitem [{FN9()}]{FN9}%
  \BibitemOpen
  \href@noop {} {}\bibinfo {note} {While the strength of the SHG in the
  tight-binding model is slightly smaller than in the Weyl Hamiltonian due to
  the finite temperature effect, the strength is also at the same order of
  magnitude as the analytic result.}\BibitemShut {Stop}%
\bibitem [{FN8()}]{FN8}%
  \BibitemOpen
  \href@noop {} {}\bibinfo {note} {The JDOS is defined
  as$\rho(\varepsilon)\equiv 2/(8\pi^3) \int d \bm k \delta(\varepsilon_c(\bm
  k) - \varepsilon_v(\bm k) - \varepsilon)$ where $\varepsilon_{c(v)}(\bm k)$
  denotes the conduction (valence) band dispersion.}\BibitemShut {Stop}%
\end{thebibliography}

\clearpage


\appendix
\renewcommand{\theequation}{S\arabic{equation}}
\setcounter{equation}{0}
\renewcommand{\thefigure}{S\arabic{figure}}
\setcounter{figure}{0}

\begin{widetext}
\begin{center}
\textbf{\large Supplemental Material: ``Current-induced second harmonic generation in  inversion-symmetric Dirac and Weyl semimetals"}
\end{center}

\section*{S1. Derivation of the response tensor from Weyl Hamiltonian}
In this section, we calculate the SHG response tensor under finite current in $z$-direction using the Weyl Hamiltonian $H_\mathrm{Weyl} = \chi \bm d(\bm k) \cdot \bm \sigma - \mu \sigma_0$ where $\bm d (\bm k) = (d_x, d_y, d_z) = ta(k_x-k_{0x}, k_y-k_{0y}, k_z-k_{0z})$ and its eigenvectors $\ket{0}$ and $\ket{1}$ satisfy $H_\mathrm{Weyl} \ket{1}=\varepsilon_1 \ket{1}= (\chi d - \mu) \ket{1}$ and $H_\mathrm{Weyl} \ket{0} =\varepsilon_0\ket{0}= (- \chi d-\mu) \ket{0}$ with $d=|\bm d (\bm k)|$. We denote the response tensor from a single Weyl point as $\sigma^{abc}_\mathrm{single}(\omega)$ where the indices $a, b$ and $c$ run over $\{ x, y, z \}$. This consists of four parts $\sigma^{abc}_\mathrm{2p, I}, \sigma^{abc}_\mathrm{2p, II}, \sigma^{abc}_\mathrm{1p, I}$, and $\sigma^{abc}_\mathrm{1p, II}$ as shown in Eqs.~(\ref{eq:SHG})-(\ref{eq:SHG1pII}) in the main text. Since $w^{ab} = 0$ for the Weyl Hamiltonian, $\sigma^{zzz}_\mathrm{1p, I}$ and $\sigma^{zzz}_\mathrm{2p, I}$ are zero. Taking a sum for the band indices and assuming $\omega>0$, we obtain the simpler form of $\sigma^{abc}_\mathrm{2p, II}$ and $\sigma^{abc}_\mathrm{1p, II}$. For $\chi=+1$, they are written as
\begin{align}
    \sigma^{abc}_\mathrm{2p, II}(\omega)&=\frac{e^3}{2(\hbar \omega)^2}\int [d \bm k] \frac{2v^a_{01}(\{v^b_{10}, v^c_{00}\}-\{v^b_{11}, v^c_{10}\})}{\omega_{10}}f_{01} R_{\gamma}(2\omega-\omega_{10}), \label{eq:SHG2pIIabc_supp} \\
    \sigma^{abc}_\mathrm{1p, II}(\omega)&= - \frac{1}{8}\sigma^{abc}_\mathrm{2p, II}\left(\frac{\omega}{2}\right) + \frac{e^3}{2(\hbar \omega)^2}\int [d \bm k] \frac{v^a_{00}\{v^b_{01}, v^c_{10}\}-v^a_{11}\{v^b_{10}, v^c_{01}\}}{2\omega_{10}}f_{01} R_{\gamma}(\omega-\omega_{10}).\label{eq:SHG1pIIabc_supp} 
\end{align}
For $\chi=-1$, they are given as
\begin{align}
    \sigma^{abc}_\mathrm{2p, II}(\omega)&=\frac{e^3}{2(\hbar \omega)^2}\int [d \bm k] \frac{2v^a_{10}(\{v^b_{01}, v^c_{11}\}-\{v^b_{00}, v^c_{01}\})}{\omega_{01}}f_{10} R_{\gamma}(2\omega-\omega_{01}), \\
    \sigma^{abc}_\mathrm{1p, II}(\omega)&= - \frac{1}{8}\sigma^{abc}_\mathrm{2p, II}\left(\frac{\omega}{2}\right) + \frac{e^3}{2(\hbar \omega)^2}\int [d \bm k] \frac{v^a_{11}\{v^b_{10}, v^c_{01}\}-v^a_{00}\{v^b_{01}, v^c_{10}\}}{2\omega_{01}}f_{10} R_{\gamma}(\omega-\omega_{01}).
\end{align}
For the definitions of $v^a_{mn}$, $\{ v^a_{mn}, v^b_{np} \}$, $\omega_{mn}$, $f_{mn}$, and $R_\gamma(x)$, see the main text. By a straightforward calculation, we can check that the final results do not depend on the chirality $\chi$ and thus we assume $\chi = +1$ below.

In the following, we evaluate the quantities given by Eqs.~(\ref{eq:SHG2pIIabc_supp}) and (\ref{eq:SHG1pIIabc_supp}). First, due to the symmetry, it turns out that non-zero independent components of the tensor are only $zzz$-, $zxx$- and $xzx$-components. To calculate them, we use 
\begin{align*}
     v^x_{01}&=\chi v \bra{0}\sigma_x\ket{1}= \chi v\left(\frac{d_x}{\sqrt{d_x^2+d_y^2}}\frac{d_z}{d}-i \frac{d_y}{\sqrt{d_x^2+d_y^2}}\right), \qquad v^x_{10}=(v^x_{01})^*,\\
    v^x_{00}&=\chi v \bra{0}\sigma_x\ket{0}= -\frac{\chi v d_x}{d}, \quad v^x_{11}= \chi v \bra{1}\sigma_x\ket{1}=  \frac{\chi v d_x}{d},\\
    v^z_{01}&=\chi v \bra{0}\sigma_z\ket{1}= - \frac{\chi v\sqrt{d_x^2+d_y^2}}{d}, \quad v^z_{10}= v^z_{01}, \quad
     v^z_{00}=\chi v \bra{0}\sigma_z\ket{0}= -\frac{\chi v d_z}{d}, \quad v^z_{11}= \chi v \bra{1}\sigma_z\ket{1}=  \frac{\chi v d_z}{d}.
\end{align*}
The distribution function under the electric field in $z$-direction is given as $f_n = f^{(0)}_n + \calE_z \frac{\partial f^{(0)}_n}{\partial k_z} +  \calE_z^2  \frac{\partial^2 f^{(0)}_n}{\partial k_z^2} + \cdots$ where $\calE_z=e\tau E/\hbar$ and we truncate the distribution function truncated up to the second order of $\calE_z$. Using this form of the distribution function, the terms of the 0-th and the 2-nd order of $\calE_z$ vanishes due to the symmetry and only the 1-st order term remains. Therefore, the distribution functions in Eqs.~(\ref{eq:SHG2pIIabc_supp}) and (\ref{eq:SHG1pIIabc_supp}) are replaced as $f_{01} \to \calE_z \frac{\partial f^{(0)}_{01}}{\partial k_z}$. The derivative $\frac{\partial f^{(0)}_{01}}{\partial k_z}$ is calculated as 
\begin{align*}
     \frac{\partial f^{(0)}_{01}}{\partial k_z}&=-\delta(d-|\mu|)\frac{\chi v \hbar d_z}{d},
\end{align*}
where $\mu\neq0$ and the temperature is zero. 
To clarify the calculation process, we consider the most simple one, the $zzz$-component, and show the calculation explicitly. Using the above results, $\sigma^{zzz}_\mathrm{2p, II}$ is written down as
\begin{align}
    \sigma^{zzz}_\mathrm{2p, II}(\omega)&= \calE_z \frac{e^3}{2(\hbar \omega)^2} \int [d \bm k]  \frac{4v^z_{01} v^z_{10} (v^z_{00}-v^z_{11})}{ \omega_{10}}\frac{\partial f^{(0)}_{01}}{\partial k_z} R_\gamma (2\omega-\omega_{10}) \nonumber\\
    &=\calE_z \frac{e^3 v^3}{2(\hbar \omega)^2}  \int [d \bm  k]  \frac{4(d_x^2+d_y^2)}{d^2} \frac{(- 2d_z/d)}{2d/\hbar }\delta(d-|\mu|)\frac{v \hbar d_z}{d}R_\gamma (2\omega-2d/\hbar) \nonumber\\
    &=-\calE_z \frac{2e^3 v^4}{\omega^2}  \int [d \bm  k]  \frac{d^2_z(d_x^2+d_y^2)}{d^5} \delta(d-|\mu|)R_\gamma (2\omega-2d/\hbar) \nonumber\\
    &\simeq-\frac{e\tau E}{\hbar} e^3v^4\hbar^2\int \frac{d \bm  k}{(2\pi)^3}  \delta(d-|\mu|)\frac{d^2_z(d_x^2+d_y^2)}{d^7} \frac{1}{ 2\omega-2d/\hbar-i\gamma} \label{eq:dimensionful}
\end{align}
Here, we assume that $\gamma$ is sufficiently small that the resonant factor $R_\gamma(x)$ behaves like $\delta(x)$ and taking the factor $1/\omega^2$ into the integrand in the last line of Eq.(\ref{eq:dimensionful}). To clarify the physical dimension, we didimentionalize several quantities in Eq.~(\ref{eq:dimensionful}). For this purpose, we factorize $v$ as $v=ta/\hbar$ where $t$ and $a$ are constants with dimension of energy and length respectively. In lattice models, $t$ and $a$ correspond to the hopping amplitude and the lattice constant. Using these quantities, we obtain
\begin{align}
    \sigma^{zzz}_\mathrm{2p, II}(\omega)&=-\frac{e\tau E }{\hbar} \frac{e^3 t^4 a^4}{\hbar^2}\int \frac{d \bm  k}{(2\pi)^3}  \delta(d-|\mu|)\frac{d^2_z(d_x^2+d_y^2)}{d^7} \frac{1}{ 2\omega-2d/\hbar-i\gamma}  \nonumber\\
    &=-\frac{e\tau E a}{\hbar} \frac{e^3 t^4}{\hbar^2}\int \frac{(a^3 d \bm  k)}{(2\pi)^3}  \frac{1}{t}\delta(d/t-|\mu|/t)\frac{1}{t^3}\frac{(d_z/t)^2\{(d_x/t)^2+(d_y/t)^2\}}{(d/t)^7} \frac{\hbar/t}{ \hbar\omega/t-d/t-i\hbar\gamma/(2t)} \nonumber\\
    &=-\frac{e\tau E a}{\hbar} \frac{\hbar}{t} \frac{e^3 }{h^2}\int \frac{dx dy dz}{2\pi}  \delta(D-|M|)\frac{D_z^2(D_x^2+D_y^2)}{D^7} \frac{1}{ \Omega-D-i\Gamma/2}\nonumber\\
    &\equiv -\frac{e\tau E a}{\hbar} \frac{\hbar}{t} \frac{e^3 }{h^2} I(\Omega; M, \Gamma/2), \label{eq:dimensionless}
\end{align}
with $\alpha = a k_\alpha, D_\alpha = d_\alpha/t~(\alpha=x, y, z), D=d/t, M=\mu/t, \Omega=\hbar \omega/t$ and $ \Gamma=\hbar \gamma/t$. Then, the problem is reduced to evaluate the integral $I(\Omega; M, \Gamma)$, which is defined as
\begin{align*}
    I(\Omega; M, \Gamma) = \int \frac{dxdydz}{2\pi} \delta(D-|M|)  \frac{(z - z_0)^2 \{(x-x_0)^2+(y-y_0)^2\}}{D^7}\frac{1}{\Omega-D-i\Gamma},
\end{align*}
with $\alpha_0 = a k_{0\alpha}~(\alpha=x, y, z)$. To evaluate it, we change the variables as $x=r \sin \theta \cos \phi + x_0$, $y=r \sin \theta \sin \phi + y_0$, and $z=r \cos \theta + z_0$. Then, we can perform the integration as
\begin{align}
    I(\Omega; M, \Gamma) &=  \int_0^{\rho_c} dr \int_0^\pi d\theta \int_0^{2\pi} \frac{d\phi}{2\pi}(r^2 \sin \theta) \delta(r-|M|)  \frac{(r^2 \cos^2 \theta) ( r^2 \sin^2 \theta)}{r^7}\frac{1}{\Omega-r-i\Gamma} \nonumber\\
     &=  \left( \int_0^\pi d\theta \cos^2 \theta \sin^3 \theta \right)  \frac{|M|^{-1}}{\Omega-|M|-i\Gamma}\nonumber \\
     &=  \frac{4}{15} \frac{|M|^{-1}}{\Omega-|M|-i\Gamma}. \label{eq:integral}
\end{align}
Applying Eq.~(\ref{eq:integral}) to Eq.~(\ref{eq:dimensionless}), we obtain
\begin{align}
    \sigma^{zzz}_\mathrm{2p, II}(\omega)&=-\frac{4}{15}\frac{e\tau E a}{\hbar} \frac{\hbar}{t} \frac{e^3}{h^2}  \frac{t}{|\mu|}\frac{t/\hbar}{\omega-|\mu/\hbar|-i \gamma / 2}. \label{eq:2pIIzzz_suppl}
\end{align}
For the one-photon contribution $\sigma^{zzz}_\mathrm{1p, II}$, 
it turns out that $\sigma^{zzz}_\mathrm{1p, II}(\omega)=-\sigma^{zzz}_\mathrm{2p, II}(\omega/2)/16$ 
because the second term of Eq.~(\ref{eq:SHG1pIIabc_supp}) is equal to $\sigma^{zzz}_\mathrm{2p, II}(\omega/2)/16$. Thus, we obtain \begin{align}
   \sigma^{zzz}_\mathrm{1p, II}(\omega)&=\frac{1}{16}\frac{4}{15}\frac{e\tau E a}{\hbar} \frac{\hbar}{t} \frac{e^3}{h^2}  \frac{t}{|\mu|}\frac{t/\hbar}{\omega/2-|\mu/\hbar|-i \gamma / 2} \nonumber\\
   &=\frac{1}{30}\frac{e\tau E a}{\hbar} \frac{\hbar}{t} \frac{e^3}{h^2}  \frac{t}{|\mu|}\frac{t/\hbar}{\omega-2|\mu/\hbar|-i \gamma}. \label{eq:1pIIzzz_suppl}
\end{align}
This simplification only occurs for the $zzz$-component. For the $zxx$- and $xzx$-components, we have to calculate the one-photon contribution itself respectively.
Then, we finally obtain $\sigma^{zzz}_\mathrm{single}$, the sum of the one-photon contribution [Eq.~(\ref{eq:2pIIzzz_suppl})] and the two-photon contribution [Eq.~(\ref{eq:1pIIzzz_suppl})], given as Eq.~(\ref{eq:SHGWeylHamTot}) in the main text.

The other components, the $zxx$- and $xzx$-components, are also calculated in the same manner. For the Weyl Hamiltonian, the $abb$-component and the $aba$-component are the sum of the two-photon ($\sigma^{abc}_\mathrm{2p,II}$) and one-photon ($\sigma^{abc}_\mathrm{1p,II}$) components and each component is given as
\begin{align}
\sigma^{abb}_\mathrm{2p,II}(\omega)&=\frac{e^3}{2\hbar^2 \omega^2}\int [d \bm k]\frac{4 v^a_{01} v^b_{10} (v^b_{00}-v^b_{11}) }{\omega_{10}}f_{01} R_\gamma (2\omega-\omega_{10}),\\
\sigma^{abb}_\mathrm{1p,II}(\omega)
&= -\frac{1}{8} \sigma^{abb}_\mathrm{2p, II}(\omega/2) + \frac{e^3}{2\hbar^2 \omega^2}\int [d \bm k] \frac{v^b_{01} v^b_{10} (v^a_{00}-v^a_{11}) }{\omega_{10}} f_{01} R_\gamma (\omega-\omega_{10}),\\
\sigma^{aba}_\mathrm{2p,II}(\omega)&=\frac{e^3}{2\hbar^2 \omega^2}\int [d \bm k]\frac{2 v^a_{01} \{ v^b_{10} (v^a_{00}-v^a_{11}) + v^a_{10} (v^b_{00}-v^b_{11}) \} }{\omega_{10}}f_{01} R_\gamma (2\omega-\omega_{10}),\\
\sigma^{aba}_\mathrm{1p,II}(\omega)
&= -\frac{1}{8} \sigma^{aba}_\mathrm{2p, II}(\omega/2) + \frac{e^3}{2\hbar^2 \omega^2}\int [d \bm k]  \frac{(v^a_{01}v^b_{10}+v^b_{01}v^a_{10})(v^a_{00}-v^a_{11})}{2\omega_{10}} f_{01} R_\gamma (\omega-\omega_{10}).
\end{align}
Performing the straightforward calculation, we obtain
\begin{align}
    \sigma^{zxx}_\mathrm{2p, II}(\omega)&=-\frac{2}{15}\frac{e\tau E a}{\hbar} \frac{\hbar}{t} \frac{e^3}{h^2}  \frac{t}{|\mu|}\frac{t/\hbar}{\omega-|\mu/\hbar|-i \gamma / 2}, \\
    \sigma^{zxx}_\mathrm{1p, II}(\omega)&=\frac{3}{10}\frac{e\tau E a}{\hbar} \frac{\hbar}{t} \frac{e^3}{h^2}  \frac{t}{|\mu|}\frac{t/\hbar}{\omega-2|\mu/\hbar|-i \gamma }, \\
    \sigma^{xzx}_\mathrm{2p, II}(\omega)&=\frac{1}{5}\frac{e\tau E a}{\hbar} \frac{\hbar}{t} \frac{e^3}{h^2}  \frac{t}{|\mu|}\frac{t/\hbar}{\omega-|\mu/\hbar|-i \gamma / 2}, \\
    \sigma^{xzx}_\mathrm{1p, II}(\omega)&=-\frac{7}{60}\frac{e\tau E a}{\hbar} \frac{\hbar}{t} \frac{e^3}{h^2}  \frac{t}{|\mu|}\frac{t/\hbar}{\omega-2|\mu/\hbar|-i \gamma}.
\end{align}

For comparison purpose, we also calculate the SHG response tensor $\sigma^{xxx}_\mathrm{2D}$ of two-dimensional Dirac Hamiltonian under a DC electric field $\bm E_\mathrm{DC}= (E, 0, 0)$. This has been studied in the previous studies~\cite{Cheng2014}. The Hamiltonian is
\begin{align}
    H_\mathrm{2D}&= \chi v \bm p \cdot \bm \sigma - \mu \sigma_0, \label{eq:2D_Dirac}
\end{align}
where $\chi = \pm 1$, $v=ta/\hbar$, $\bm p = (\hbar k_x, \hbar k_y, 0)$. For simplicity, we only consider one component $\sigma^{xxx}$ because the difference from other components is only the numerical factor. Since the Hamiltonian (\ref{eq:2D_Dirac}) is linear in momentum and a two-band model, the SHG response tensor is simplified as
\begin{align}
    \sigma^{xxx}_\mathrm{2D} = \sigma(\omega) - \frac{1}{16} \sigma(\omega/2).
\end{align}
Here, the first (second) term corresponds to the two-photon (one-photon) contribution defined by Eq.~(\ref{eq:SHG2pII}) (Eq.~(\ref{eq:SHG1pII})) in the main text. Using Eq.~(\ref{eq:SHG2pII}), $\sigma(\omega)$ is calculated as
\begin{align}
    \sigma(\omega)&= \frac{e \tau E}{\hbar}  \frac{e^3}{2(\hbar \omega)^2} \int [d \bm k] \frac{4v^x_{mn} v^x_{np} v^x_{pm}}{\omega_{mp}+\omega_{np}} \frac{\partial f_{mn}}{\partial k_x} R_\gamma (2 \omega - \omega_{nm})
    \\
    &=-\frac{e \tau E}{\hbar}  \frac{e^3}{2(\hbar \omega)^2} \int [d \bm k] \frac{4v^x_{01} v^x_{10} (v^x_{11}-v^x_{00})}{\omega_{10}} \frac{\partial f_{01}}{\partial k_x} R_\gamma (2 \omega - \omega_{10}).
\end{align}
Using these relations
\begin{align}
    v^x_{01}=-i \frac{ta}{\hbar} \frac{d_y}{d}, \qquad v^x_{10}= (v^x_{01})^*, \qquad v^x_{11} - v^x_{00}=\frac{2ta}{\hbar}\frac{d_x}{d},
\end{align}
we obtain
\begin{align}
\sigma(\omega)
    =- \frac{e \tau E a}{\hbar} \frac{e^3}{h^2} \frac{a\hbar}{t} \frac{\pi}{4}    \frac{t^2}{|\mu|^2} \frac{t/\hbar}{\omega-|\mu/\hbar|- i\gamma/2}.
\end{align}
Then, the result is
\begin{align}
    \sigma^{xxx}_\mathrm{2D}(\omega) &= \frac{e\tau E a}{\hbar} \frac{a\hbar}{t} \frac{e^3}{h^2} \left\{G_\mathrm{2p}(\omega)+G_\mathrm{1p}(\omega)\right\},
\end{align}
with
\begin{align}
    G_\mathrm{2p}(\omega)=- \frac{\pi}{4} \frac{t^2}{|\mu|^2}\frac{t/\hbar}{\omega - |\mu/\hbar| - i \gamma/2}, \\
    G_\mathrm{1p}(\omega)=\frac{\pi}{32}
    \frac{t^2}{|\mu|^2}\frac{t/\hbar}{\omega - 2|\mu/\hbar| - i \gamma}. 
\end{align}

\newpage

\section*{S2. Deviation of the peaks in SHG and JDOS spectra}

In this section, we explain in detail why the positions of the peak around $\hbar \omega \sim 1.1$ in the SHG and the peak of the van Hove singularity (vHs) in the JDOS spectra are different. This difference is seen in Figs.~(\ref{Fig:TB})~(b) and (d, e) in the main text. Figs.~(\ref{Fig:S2})~(a) and (b) show the difference more clearly. Note that the half of the frequency in the JDOS is used to compare it with the frequency of the SHG spectrum because the main contribution of the SHG is the two-photon resonance term Eq.~(\ref{eq:SHG2pII}) in the main text. In Fig.~(\ref{Fig:S2})~(a), only the contribution of Eq.~(\ref{eq:SHG2pII}) is shown.

As we mentioned in the main text, the reason is that the group velocity vanishes at the peak in the JDOS spectrum. Since the SHG response tensor contains the group velocity [See Eq.~(\ref{eq:SHG}) in the main text], the response tensor at the vHs point also vanishes and the peak in the SHG spectrum slightly shifts from the vHs point. This shift should be determined by the product of the density of states and the group velocity.

To confirm the above argument, we compare these spectra with the DC conductivity, which is a quantity similar to the product of the density of states and the group velocity. The DC conductivity is defined via $j^z(\omega=0)=\sigma^z_\mathrm{DC} E^z(\omega=0)$ and given as
\begin{align}
    \sigma^z_\mathrm{DC} = -\frac{e^2 \tau}{\hbar} \int \frac{d \bm k}{(2\pi)^3} \sum_n  \frac{\partial f^{(0)}_n}{\partial k_z} v^z_{nn}.
\end{align}
The comparison of the SHG spectrum, the JDOS spectrum and the DC conductivity is shown in Fig.~(\ref{Fig:S2}). Seeing Figs.~(\ref{Fig:S2})~(a)~and~(c), the behavior of the peak height in the SHG spectrum is similar to that of the DC conductivity. Both the peak height of the SHG and the DC conductivity develop from the vHs point (around $\hbar \omega/t, \mu/t \sim 1.0$) and reach the largest value around $\hbar \omega/t, \mu/t \sim 1.1$. This coincidence strongly supports the above argument.

\begin{figure*}[h]
\includegraphics[width=12cm]{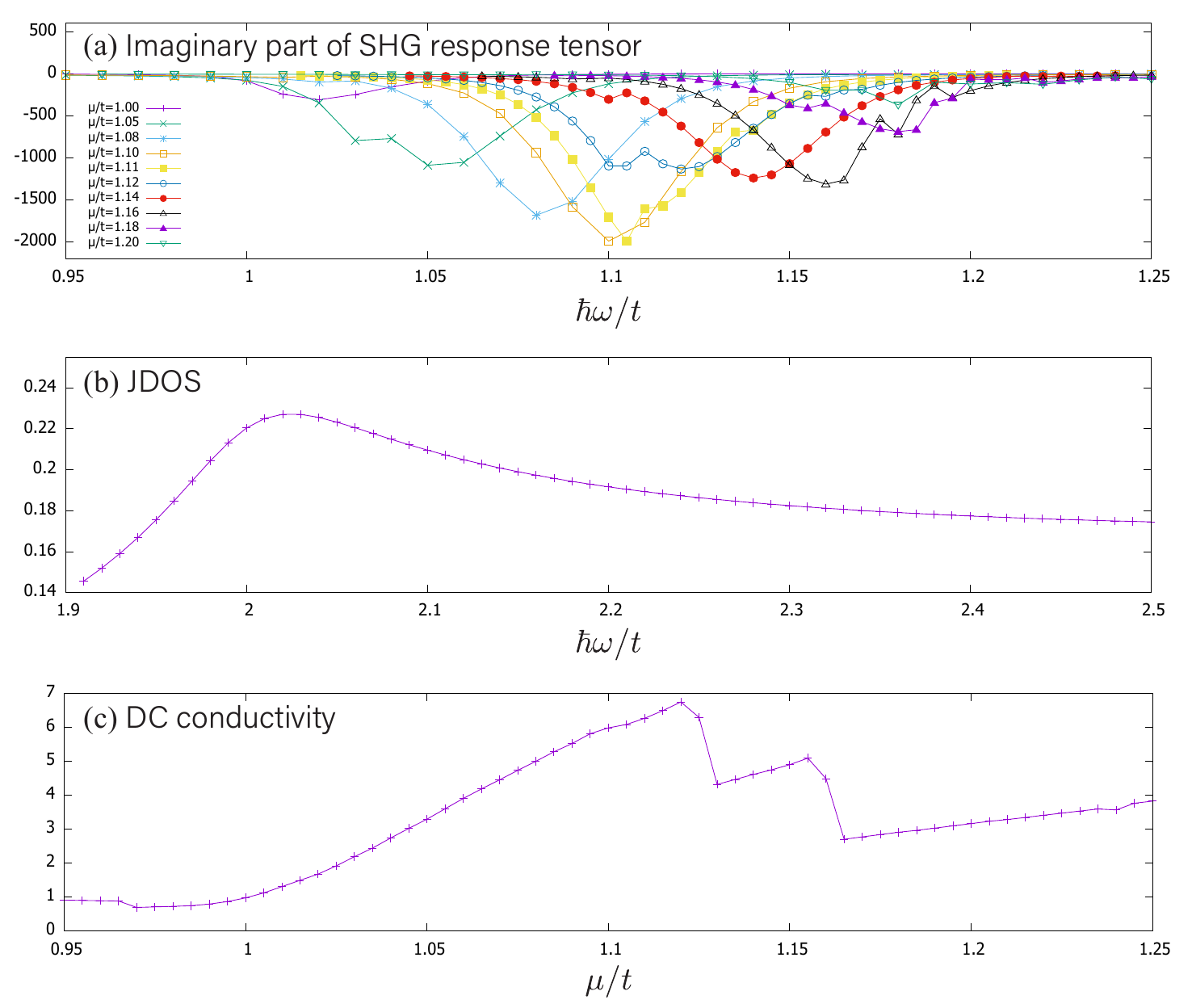}
\caption{
(a) Imaginary part of the SHG spectra with different values of the chemical potential, (b) joint density of states (JDOS), and (c) DC conductivity of the tight-binding model for Dirac semimetals~[Eq.~(\ref{eq:TBmodel})] in the main text. Note that only one contribution of two-photon resonance Eq.~(\ref{eq:SHG2pII}) is shown in Fig.~(a).
}
\label{Fig:S2}
\end{figure*}

\newpage

\section*{S3. Tight-binding calculation for Weyl semimetals}

In this section, we show the results of the tight-binding calculation for Weyl semimetals. While we show the results for Dirac semimetals in the main text, the qualitative behavior of Weyl semimetals is same as that of Dirac semimetals.

We use a tight-binding model with two bands describing Weyl semimetals $H_\mathrm{TB}=\bm d(\bm k)\cdot \bm \sigma$ with
\begin{align}
    \bm d (\bm k) &= t(\sin (a k_x) , \sin (a k_y) , 2-m/t-\cos (a k_x) -\cos (a k_y) -\cos (a k_z)), \label{eq:TBWeyl}
\end{align}
adopted from Ref.~\cite{Ramamurthy2015}. The band structure of this model is shown in Fig.~\ref{Fig:S3}~(a). We calculate the SHG response tensor $\sigma^{zzz}_\mathrm{WSM}(\omega)$ under the electric field in $z$-direction $\bm E=(0,0,E_z)$. 

The results are shown in Figs.~\ref{Fig:S3}~(b) and (c). The divergent behavior with approaching $\mu$ to zero is observed and the additional peaks around $\hbar \omega /t \sim 1.0$ are also confirmed. Therefore, the qualitative results are completely same as those of Dirac semimetals shown in the main text.
From the quantitative viewpoint, it is expected that the SHG response of Weyl SMs becomes half of that of the Dirac SMs since the number of Weyl nodes is half. However, the value of our result is smaller than  expected. We consider that this is because of the detail of the tight-binding models. To claim more quantitative argument, we need to use more similar models for Dirac and Weyl SMs.

\begin{figure*}[h]
\includegraphics[width=16cm]{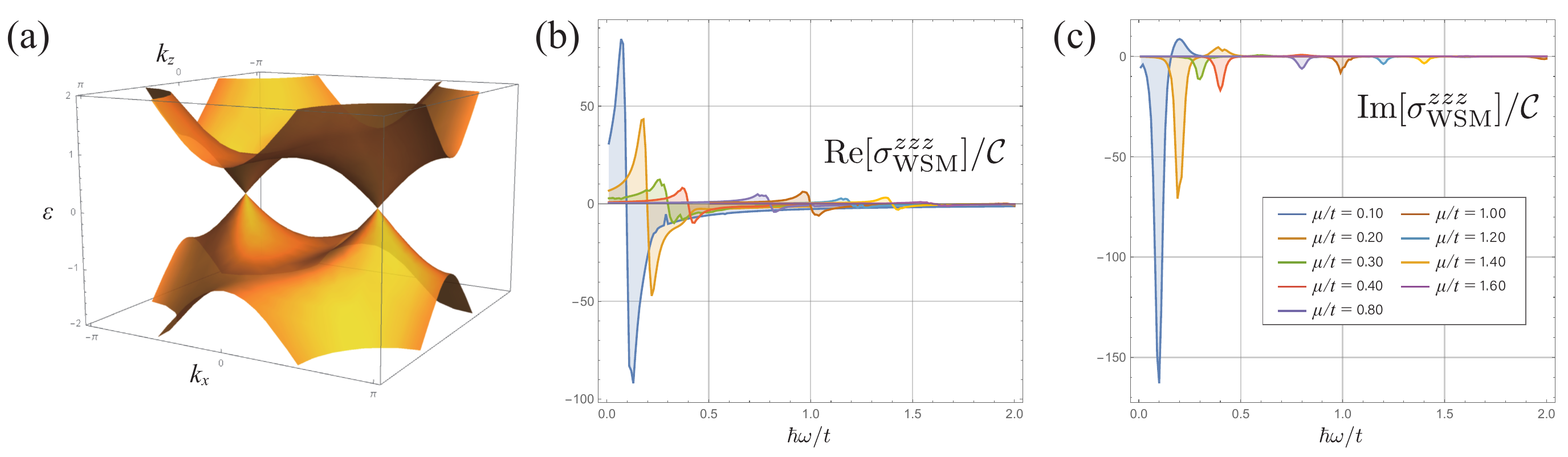}
\caption{
(a) Energy dispersion and (b, c) real and imaginary part of the CISHG response tensor $\sigma^{zzz}_\mathrm{WSM}(\omega)$ of the tight-binding model [Eq.~(\ref{eq:TBWeyl})] for Weyl semimetals. calculated with the tight-binding Hamiltonian. The values in the vertical axes are normalized by a constant $\calC=(e\tau E_z a/\hbar)\cdot(\hbar/t)\cdot(e^3/h^2)$. The parameters set as $t=1.0$, $m=0$, $\beta=100$ and $\gamma=0.01$.
}
\label{Fig:S3}
\end{figure*}

\end{widetext}
\end{document}